\documentclass[pra,showpacs,twocolumn]{revtex4}
\usepackage{epsfig}
\usepackage{array}
\usepackage{amsmath}

\begin{document}

\title{A possibility for precise Weinberg angle measurement
in centrosymmetric crystals with axis}

\author{T. N. Mukhamedjanov, O. P. Sushkov}

\affiliation{School of Physics, University of New South Wales,\\
 Sydney 2052, Australia}

\begin{abstract}
We demonstrate that parity nonconserving interaction due to the nuclear weak charge $Q_W$
leads to nonlinear magnetoelectric effect in centrosymmetric paramagnetic crystals.
It is shown that the effect exists only in crystals with special symmetry axis ${\bf k}$. 
Kinematically, the correlation (correction to energy) has the form
$H_{\textrm{PNC}} \propto Q_W {\bf E} \cdot [{\bf B}\times{\bf k}]({\bf B}\cdot{\bf k})$, 
where ${\bf B}$  and ${\bf E}$ are external magnetic and electric fields.
This gives rise to the magnetic induction
${\bf M}_{\textrm{PNC}} \propto Q_W \left\{{\bf k}({\bf B}\cdot[{\bf k}\times{\bf E}]) +
[{\bf k}\times{\bf E}]({\bf B}\cdot{\bf k})\right\}$.
To be specific, we consider rare-earth trifluorides and, in particular,
dysprosium trifluoride which looks the most suitable for experiment.
We estimate the optimal temperature for the experiment to be of a few kelvin.
For the magnetic field $B=1\ \textrm{T}$ and the electric field $E=10\ \textrm{kV/cm}$, 
the expected magnetic induction is $4 \pi M_{\textrm{PNC}} \sim 0.5 \cdot 10^{-11} \ \textrm{G}$,
six orders of magnitude larger than the best sensitivity currently
under discussion.
Dysprosium has several stable isotopes, and so, comparison of the effects for 
different isotopes provides possibility for precise measurement of the Weinberg angle.
\end{abstract}


\maketitle

\section{introduction}

Studies of atomic parity nonconservation (PNC) provide important tests of the standard model of electroweak interactions and impose stringent constraints on the new physics \cite{PDG}. 
The effects of PNC in atoms due to the nuclear weak charge $Q_W$ have been successfully measured for bismuth \cite{BaZo78}, lead \cite{Mee93}, thallium \cite{Edwa95}, and cesium \cite{WoBe97,BeWi99}. The authors of \cite{WoBe97,BeWi99} have also performed measurements of the nuclear spin-dependent PNC effect in Cs caused by the nuclear anapole moment (e.g., see text \cite{Khriplo91}), which remains the only measurement of this kind to date.

In the present work we concentrate on effects caused by the nuclear weak charge in solids.
We predict a new kind of (nonlinear) magnetoelectric effect in centrosymmetric crystals due to the violation of parity ($P$) at fundamental level, proportional to $Q_W$.
Although we show that the experimental observation of the proposed effect is possible at the current level, the uncertainty of theoretical calculation would not allow for an accurate interpretation in terms of the nuclear weak charge $Q_W$ itself.
Instead, the potential importance for fundamental physics lies in the isotope dependence
of the effect we suggest.
We argue that comparing the value of the effect in samples with different isotope
content, one can, in principle, determine the Weinberg angle with high precision.
In this respect the idea is similar to the suggestion presented in Ref.~\cite{DzuFla86} of an atomic PNC measurements in rare-earth Dy, and the later suggestion of Ref.~\cite{DMille95} for a measurement with atomic Yb.

The idea to use solids for studies of fundamental symmetry violations dates back to 1968, 
when Shapiro suggested that the existence of the permanent electric dipole moment (EDM) of electron 
should lead to linear magnetoelectric effect in solids \cite{Shapi68}.
(The existence of the permanent EDM of a quantum particle violates both parity ($P$) and time-reversal ($T$) 
symmetries, see, e.g., \cite{KhriLa97}.)
However, the results of the early experiment \cite{VaKo78} (1978) performed in Ni-Zn ferrite were not impressive due to experimental limitations.
In the recent paper \cite{Lamo02}, Lamoreaux suggested that application of novel experimental techniques in garnet materials should lead to substantial increase in sensitivity compared to the previous attempt; this suggestion renewed the interest in the original idea.
Calculations \cite{MuDzu03} have shown that improvement of the statistical sensitivity to the electron EDM of several orders of magnitude was possible compared to the current experimental limit on this value \cite{ReCo02}.
At the same time, the experimental search was initiated---the first results are already available in the literature, \cite{HeE05} (see also Ref.~\cite{LiLa04}).
The other suggestions for fundamental symmetry violation search in solids include proposed search for manifestations of the nuclear anapole moment in garnets \cite{MuSu05} (violation of parity), and the search for the nuclear Schiff moment in the ferroelectric lead titanate compound \cite{MuSu05:2} (violation of $T$ and $P$ symmetries).

One of the most important issues for any solid state experiment is certainly the existence of various systematic effects.
It is getting more and more clear that the right choice of compound is crucial.
In particular, any kind of spontaneous magnetic ordering in the crystal (ferro-, antiferro-,
or ferrimagentism) should be avoided, thus, we consider paramagnetic material---rare-earth trifluoride RF$_3$.
The PNC weak interaction of electrons with the nuclei of rare-earths gives contribution to the total energy of the crystal
\begin{equation}
E_{\textrm{PNC}} \propto Q_W({\bf E} \cdot [{\bf B}\times{\bf k}])({\bf B}\cdot{\bf k}) \ ,
\label{p5:h1}
\end{equation}
in the applied electric field ${\bf E}$ and magnetic field ${\bf B}$; vector ${\bf k}$ is directed along 
the crystal symmetry axis.
Correlation similar to Eq.~(\ref{p5:h1}) was considered previously by Bouchiat and Bouchiat \cite{BoBo01} in regard to their proposal to measure nuclear anapole moment using Cs atoms trapped in solid $^{4}$He.

The effective interaction (\ref{p5:h1}) induces the following PNC macroscopic magnetization which can be measured in a magnetometry experiment:
\begin{multline}
{\bf M} = -\frac{\partial E_{\textrm{PNC}}}{\partial {\bf B}}\propto
Q_W\Big\{{\bf k}({\bf B}\cdot[{\bf k}\times{\bf E}]) \\ +
[{\bf k}\times{\bf E}]({\bf B}\cdot{\bf k})\Big\} \ .
\end{multline}
Naively, one can expect the PNC magnetization of the form 
${\bf M} \propto [{\bf B}\times{\bf E}]$, however, this kind of magnetization does not lead to any energy shift in the applied magnetic field, and hence, does not appear.
Eq.~(\ref{p5:h1}) presents the simplest energy correlation allowed kinematically. To 
generate this correlation we need a crystal that allows nontrivial second rank material tensors, so the crystal symmetry must be lower than cubic.
In addition, we need a necessarily centrosymmetric crystal in order to avoid the imitation of the PNC effect by spontaneous parity breaking in the lattice.

Rare-earth trifluoride compounds have hexagonal structure described by $C_{3,i}$ space group and, apparently, satisfy the crystallographic criteria we set.
These compounds are widely used as laser materials, and as such, their electronic structure has been studied extensively and is rather well-understood.
As it will be seen later, DyF$_3$ is, probably, especially well-suited for the experiment because of the large value of the PNC effect in this compound. Dysprosium has also particularly large number of stable isotopes which can be used for the determination of the Weinberg angle as mentioned above.

The structure of this paper is as follows.
Section II deals with atomic estimates involved in the problem. Starting from the 
electron-nucleus weak interaction, we consider an ion in the lattice environment and derive the effective single-ion PNC Hamiltonian.
In Section III we average this Hamiltonian over the crystal structure of the compound, which gives the kinematical structure of the effect and its dependence on the external fields. 
Section IV presents our results and discussion of uncertainties involved.

\section{Single ion effective Hamiltonian}

The weak charge-induced PNC interaction between electron and nucleus is of the form 
(see, e.g., \cite{Khriplo91})
\begin{equation}
H_W = - \frac{G}{\sqrt{2}} \delta({ \bf r}) \gamma_5 \frac{Q_W}{2}\ .
\label{p5:wk1}
\end{equation}
Here $G$ is the Fermi coupling constant, $\delta({ \bf r})$ is the Dirac delta function for electrons representing the nuclear density distribution, 
$\gamma_5$ is the Dirac $\gamma$-matrix.
The nuclear weak charge is $Q_W=-N+Z(1-4\sin^2\theta_W)$ where $\theta_W$ is the Weinberg angle,
$N$ is the number of neutrons, and $Z$ is the number of protons.

Let us consider an ion embedded in crystal lattice, and let us impose
an external electric field ${\bf E}$ which shifts the ion from its equilibrium position with displacement ${\bf X}$.
The ion has the orbital angular momentum ${\bf L}$ and the spin ${\bf S}$,
in this section we assume that ${\bf L}$ and  ${\bf S}$ are decoupled from each other.
Certainly, in reality ${\bf L}$ and ${\bf S}$ are coupled
due to the combined action of the spin-orbit interaction and the non-central
crystal field of the lattice at the ion site. We will consider these effects in the next section, but for now we switch these interactions off.
The interaction (\ref{p5:wk1}) is a $T$-even pseudoscalar, therefore, the effective Hamiltonian
which arises after calculation of the electronic matrix element of (\ref{p5:wk1}) must be of the form
\begin{equation}
H_{\textrm{PNC}} \propto {\bf E} \cdot [ {\bf L} \times {\bf S}] \ .
\label{p5:kin1}
\end{equation}
This is the only structure allowed kinematically, and the present section is devoted to the calculation of this effective Hamiltonian.

\begin{figure}
\begin{center}
\epsfig{figure=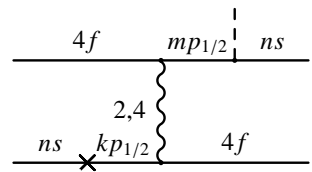,width=4cm,clip=}
\end{center}
\caption{A typical atomic perturbation theory diagram responsible for the PNC Hamiltonian (\ref{p5:kin1}). The wavy line on the diagram represents the 
residual Coulomb interaction between electrons, the weak interaction (\ref{p5:wk1}) is shown by the cross on the electron line, and the dashed line denotes interaction of electron with the external electric field $\bf E$.}
\label{p5:diag}
\end{figure}

Rare-earth trifluorides RF$_3$ are ionic crystals containing triply ionized rare-earth ions R$^{3+}$.
The general electronic configuration of these ions is $1s^2...5s^25p^64f^n$ with the number of $4f$-electrons ranging 
from 1 in Ce$^{3+}$ to 13 in Yb$^{3+}$. So, $\bf L$ and $\bf S$ are carried by the
$4f$-electrons which constitute the only open shell. $f$-electrons practically do not penetrate to the nucleus and
hence, practically do not interact with the weak charge directly. Instead, the effective interaction
arises in the third order of perturbation theory due to many-body effects,
a typical diagram is shown in Fig.~\ref{p5:diag}.
The vertical wavy line on the diagram connecting the two electron lines represents the 
residual Coulomb interaction between electrons; the weak interaction (\ref{p5:wk1}) is shown by 
the cross on the electron line; the dashed line denotes interaction of electron with the external electric field $\bf E$.
This is but one of a number of diagrams responsible for the effect, in order
to find the effective interaction (\ref{p5:kin1}) one has to evaluate all of the diagrams.
A similar atomic calculation has been performed previously in the paper~\cite{MuSu05} in regard to 
the interaction of atomic electrons with the nuclear anapole moment.
The effective interaction with the anapole moment appears in the third order of perturbation 
theory as well, and, although it has different kinematical structure, the part involving atomic 
radial integrals is practically identical to the present case.
The main difference is that the weak charge matrix element $\langle H_W\rangle$, Eq.~(\ref{p5:wk1}), shown by the
cross in the diagram Fig.~\ref{p5:diag}, has to be replaced with the matrix element of the anapole interaction $\langle H_a\rangle$.
Therefore, to estimate the effective Hamiltonian (\ref{p5:kin1}), it is not necessary to repeat the atomic calculations---it is sufficient to rescale the results of Ref.~\cite{MuSu05}. 
Analytical expressions for both the weak charge $\langle H_W\rangle$ and the anapole moment $\langle H_a\rangle$ interaction matrix elements based on the semiclassical approximation for electron wave functions can be found in the book \cite{Khriplo91}, and lead to the following ratio:
\begin{equation}
\frac{\langle s_{1/2}|H_{W}| p_{1/2}\rangle}{\langle s_{1/2}|H_a| p_{1/2}\rangle} \sim 
\frac{Q_W}{\kappa_a}\ .
\label{p5:resc}
\end{equation}
Here $\kappa_a$ is the dimensionless constant characterizing the nuclear anapole moment, it is of the order $\kappa_a \sim 0.4\,$. Rescaling the result of Ref.~\cite{MuSu05}, we find the the single-electron effective Hamiltonian for $4f$-electrons:
\begin{eqnarray}
&H_{\textrm{PNC}} \sim 2\cdot10^{-16}\ ({\bf X} \cdot [ {\bf l} \times {\bf s}]) Q_W E_0\ .
\label{p5:se}
\end{eqnarray}
Here  ${\bf l}$ and  ${\bf s}$ are the single-electron orbital angular momentum and spin, and
$E_0 = 27.2 \ \textrm{eV}$ is the atomic energy unit. The ionic displacement $\bf X$ 
is measured in units of the Bohr radius $a_B$.

Simple electrostatic considerations allow to relate the displacement $\bf X$ of the triply ionized rare-earth ion in the lattice to the strength $\bf E$ of the external electric field:
\begin{eqnarray}
{\bf P} &=& 3e\,n{\bf X}a_B = \frac{\epsilon -1}{4\pi} {\bf E}\ ;\\
{\bf X} &=& \frac{\epsilon -1}{4\pi} \frac{\bf E}{3e\,na_B}\ . \label{p5:XE}
\end{eqnarray}
Here $\bf P$ is the dielectric polarization, $e=|e|$ is the elementary charge, $n\approx 1.8 \cdot 10^{22} \ \textrm{cm}^{-3}$ is the number density of rare-earth ions in the compound, and $\epsilon \approx 14$ is the dielectric constant.
Evaluating Eq.~(\ref{p5:XE}) we get
\begin{equation}
{\bf X} [a_B] = 2.5 \cdot 10^{-8} \ {\bf E} [\textrm{V/cm}]\ .
\label{p5:X}
\end{equation}

Finally, we have to average the single-electron effective Hamiltonian (\ref{p5:se}) over the ground
state of the $4f^n$ electronic configuration of the R$^{3+}$ ion with the total momenta ${\bf L}={\bf l}_1+...+{\bf l}_n$ and ${\bf S}={\bf s}_1+...+{\bf s}_n$ (see also the remark \cite{com1}).
The averaging was performed with the help of the usual fractional parentage coefficients, see, e.g., \cite{CFP}.
The result reads:
\begin{equation}
H_{\textrm{PNC}} \sim A\ \cdot 10^{-24} \ {\bf E} \cdot [ {\bf L} \times {\bf S}]\, Q_W\, E_0\ ,
\label{p5:Hat}
\end{equation}
where the electric field ${\bf E}$ is measured in units of V/cm.
Values of the coefficient $A$ in the above equation for different rare-earth ions are listed in Table~\ref{p5:table1}. 
Note, that the effect vanishes for Gd$^{3+}$ ion which has the ground state with $L=0$.

\begin{table}[tb!]
\begin{tabular}{lrrrrrrr}
Kramers ions \\
\hline \hline
&Ce  & Nd & Sm & Gd & Dy & Er & Yb \\ \hline
$L$& 3 & 6  & 5  & 0  & 5  & 6  & 3 \\
$S$&1/2& 3/2& 5/2& 7/2& 5/2& 3/2& 1/2 \\
$A$ &$\quad 5.0 $  & $\quad 1.7 $  & $\quad 1.0 $  & $0$ & $ -1.0 $  & $ -1.7 $  & $ -5.0 $ \\
\hline
\\[-0.1cm]
non-Kramers ions\\
\hline\hline
 &Pr& Pm & Eu & Tb& Ho & Tm & \\ \hline
$L$& 5&  6 & 3  & 3 & 6  &  5 & \\
$S$& 1&  2 & 3  & 3 & 2  &  1 & \\
$A$&$\quad 2.5 $  & $\quad 1.3 $  & $\quad 0.8 $  & $-0.8$  & $ -1.3 $  & $ -2.5 $ & \\
\hline
\end{tabular} 
\caption{Values of the coefficient $A$ in the effective parity nonconserving Hamiltonian (\ref{p5:Hat}). For quick reference, we also present the values of the orbital momentum quantum number $L$ and the spin $S$ for each rare-earth ion.}
\label{p5:table1}
\end{table}

Eqs.~(\ref{p5:se}), (\ref{p5:Hat}) were obtained by rescaling the results of our previous calculation \cite{MuSu05}, where the rare-earth ions were considered in the garnet environment consisting of eight 
oxygen O$^{2-}$ ions. 
In the present case the lattice structure is different, and the environment is composed of fluorine ions F$^{-}$.
For the present calculation which concerns the Hamiltonian (\ref{p5:Hat}), however, only the spherically symmetric part of the electron density from the neighboring ions plays role. Since the ions O$^{2-}$ and F$^{-}$ have the same electronic configuration, $2p^6$, and the interatomic separation in both compounds is roughly the same, we believe the rescaling procedure gives reliable results.
At most, we can imagine an error in the coefficient $A$, Eq.~(\ref{p5:Hat}), of a factor $2-3$, which is acceptable for the purposes of the present estimate.
In the next Section, when we consider coupling to the external magnetic field, 
the anisotropic nature of the environment is very important. There, we treat the ionic environment more accurately in the framework of the Coulomb crystal field model.

\section{crystal structure and induced magnetization}

In the present section we take calculation of the PNC interaction Hamiltonian (\ref{p5:Hat}) one step further. The idea is that the PNC part is exceptionally small compared to the other terms in the full Hamiltonian for the ion in external magnetic field:
\begin{equation}
H = A_{\textrm{\it LS}}\, ({\bf L}\cdot{\bf S}) + H_{\textrm{cf}} + 
\mu_B\, {\bf B} \cdot ({\bf L} + 2{\bf S}) + H_{\textrm{PNC}}\ .
\label{p5:Hkin}
\end{equation}
(Here $A_{\textrm{\it LS}}$ is the constant for fine-structure splitting, $H_{\textrm{cf}}$ denotes the 
crystal field interaction, $\mu_B$ is the Bohr magneton.) So, it is sufficient to consider the $H_{\textrm{PNC}}$ in the first order of perturbation theory only,
\begin{equation}
\Delta E_{\textrm{PNC}} = \langle g | H_{\textrm{PNC}} | g \rangle\ .
\label{p5:De}
\end{equation}
The ground state $|g \rangle$ is determined by the first three terms in the Hamiltonian (\ref{p5:Hkin}).

For the case of isotropic environment, the direction of $\langle  {\bf L} \rangle$ and $\langle  {\bf S} \rangle$ is completely determined by the external magnetic field: $\langle  {\bf L} \rangle,\langle  {\bf S} \rangle \propto {\bf B}$; the spin-orbit coupling conserves ${\bf J} = {\bf L} + {\bf S}$, so that $\langle {\bf L} \rangle,\langle  {\bf S} \rangle \propto \langle {\bf J} \rangle$, and $\langle {\bf J} \rangle \propto {\bf B}$. Intuitively, it is then clear that the expression $\langle  {\bf L} \times {\bf S} \rangle$ in (\ref{p5:Hat}) must vanish; the result is rather easy to prove rigorously if one considers the ground state $|g\rangle=|J,J_z\rangle$.
So, the PNC effect (\ref{p5:Hat}) vanishes for environments with spherical symmetry, and the same result is expected for cubic lattices.
The situation is different, however, for an ion in the crystal environment with an axis ${\bf k}$.
This is especially obvious in the case of zero
spin-orbit coupling when $\langle  {\bf L} \times {\bf S}\rangle=
\langle  {\bf L}\rangle \times \langle{\bf S}\rangle$.
Indeed, the direction of $\langle{\bf L}\rangle$ is then determined not only by 
the external magnetic field ${\bf B}$, but also by the lattice crystal field:
in the simplest case we can write $\langle{\bf L}\rangle \propto {\bf B} + ({\bf B}\cdot{\bf k})\,{\bf k}$. The total spin $\langle{\bf S}\rangle$ is oriented along the magnetic field. Then, the value of the cross product $\langle{\bf L} \times {\bf S}\rangle \propto ({\bf B}\cdot{\bf k})[{\bf B}\times{\bf k}]$ is, generally, not zero. 
Existence of the $LS$-coupling certainly complicates the analysis, but the general picture remains the same.

The crystal field Hamiltonian $H_{\textrm{cf}}$ in Eq.~(\ref{p5:Hkin}) was chosen in the form suggested in paper~\cite{MoLe79}:
\begin{equation}
H_{\textrm{cf}} = \sum_{nm} \rho_n A_{nm} \sum_i \sqrt{\frac{4\pi}{2n+1}} Y_{nm}({\bf r}_i)\ ,
\end{equation}
where ${\bf r}_i$ is the radius-vector of the $4f$-electron, and $i$ runs over all electrons in the $4f$ shell. Parameters $A_{nm}$ do not depend on the ion, but are specific to the host compound (local environment of the ion). Together with the ion-dependent parameters $\rho_n$ they describe the effect of the host crystal lattice on the rare-earth ion.
Authors of paper \cite{MoLe79} performed consistent analysis of the optical spectra of nine different 
rare-earth ions when they dope lanthanum trifluoride by substituting lanthanum ions, and deduced 
the values of parameters $A_{nm}$ for this compound.
In our analysis, we neglect the possible difference between the local field in lanthanum 
trifluoride doped by rare-earth ions and the crystal field in generic rare-earth trifluoride compounds, and use the crystal field parameters derived in paper \cite{MoLe79}.

According to Eq.~(\ref{p5:Hat}), in order to calculate $\langle g | H_{\textrm{PNC}} | g \rangle$ we
need to find the expectation value of $[{\bf L} \times {\bf S}]$.
To be specific, let us consider non-Kramers ions, i.e., ions that contain even number of
electrons. In this case the ground state is non-degenerate and, as the direct calculation shows, the expectation value $\langle [{\bf L} \times {\bf S}] \rangle$ vanishes in the zero magnetic field.
So, $\langle H_{\textrm{PNC}} \rangle=0$, if $B = 0$, and, certainly, this agrees with the general statement that the usual weak interaction cannot
generate linear Stark effect \cite{Khriplo91}. In the nonzero magnetic field the expectation value is of the form
\begin{multline}
\langle g| [{\bf L} \times {\bf S}] | g \rangle = a[{\bf B}\times{\bf k}]({\bf B}\cdot{\bf k}) \\
+ b\left[ {\bf k} B^2 - {\bf B} ({\bf B}\cdot{\bf k})\right],
\label{p5:eff}
\end{multline}
where ${\bf k}$ is the unit vector orthogonal to the plane of hexagonal symmetry,
and $a$ and $b$ are some constants derived from calculation.
The first term in (\ref{p5:eff}) corresponds to the simple structure that follows from most general arguments, as discussed in the
beginning of this Section; a similar structure was considered in the literature \cite{BoBo01} for Cs trapped in solid $^4$He.
The nature of the second term is not as obvious.
An interesting feature of this term is that the vector $b{\bf k}$ is a pseudovector:
while a true vector is forbidden in centrosymmetric lattice, a pseudovector is
generally allowed. We have confirmed numerically that the contribution proportional to $b{\bf k}$ 
appears only due to 4th and 6th multipolarities of the crystal field---indeed, one 
can construct a 
pseudovector from the irreducible tensors of 4th and 6th rank, but not from a 2nd rank tensor. 
For example, one can construct a pseudovector from the 4th rank irreducible tensor $\hat{t}$, which represents the crystal field, in the following way:
\begin{equation}
bk_i \propto \epsilon_{ijk} \ t_{jlmn} \ t_{lmnp} \ t_{rsuv} \ t_{uvpq} \ t_{qrsk}\ .
\label{p5:t5}
\end{equation}
Here $\epsilon_{ijk}$ is the usual totally antisymmetric tensor, and summation over the repeated indices is implied.
At zero temperature, the combination (\ref{p5:t5}) does not appear explicitly in the expression
for the coefficient $b$---in this case the coefficient does not have a perturbation theory structure, and, hence,
the answer cannot be presented as a simple series in powers of crystal field.
However, we also considered a nonzero temperature case. At high temperature,
$kT \gg H_{\textrm{cf}}, A_{LS}\ $, the pseudovector $b{\bf k}$ can be represented as a series
in powers of $H_{\textrm{cf}}/kT$. In this case numerics clearly indicate that, in agreement with (\ref{p5:t5}), the leading contribution to $b{\bf k}$ appears in the fifth order in crystal field.

The system of electron spins in RF$_3$ at sufficiently low temperature undergoes transition 
to the ordered state. It is clear that the lowest temperature in the experiment is limited to 
the transition temperature. However, the existing experimental data on RF$_3$ is, unfortunately, scarce.
It is known that the electron spins in TbF$_3$ order ferromagnetically below $3.95\ \textrm{K}$ \cite{HoGu70}, 
and the spins in HoF$_3$ order antiferromagnetically at $0.53\ \textrm{K}$ \cite{BleGre88}. From the results of Ref.~\cite{LeLGa84} we also know that (Ce,Nd,Pr)F$_3$ compounds remain 
paramagnetic at least down to $2\ \textrm{K}$. It is reasonable to conclude that 
other rare-earth trifluorides have transition temperature of $\lesssim 1\ \textrm{K}$ as well, 
so, in our estimates, we assume the temperature of $1\ \textrm{K}$.
For non-Kramers ions where the ground state is not degenerate the effect is, actually,
temperature independent as soon as the temperature in the compound is above the temperature of collective spin ordering $\sim 1\ \textrm{K}$, and below the first energy interval in the crystal field splitting ($\sim10\ \textrm{K}$ for compounds where effect is the largest).
However, the temperature dependence is very important for Kramers ions where the ground state is degenerate, and the effect there scales as $1/T$. In the calculation we set $T=1\ \textrm{K}$.

\begin{table*}[tb!]
\begin{tabular}{lrlrlrlrlrlrlrl}
 \multicolumn{1}{l}{Kramers ions} \\
\hline \hline
 & Ce && Nd && Sm && Gd && Dy && Er && Yb \\
\hline
 $a$ & $0.30 $ &$ \cdot 10^{-4}$ & $0.40 $ &$ \cdot 10^{-3}$ & $\quad0.21$ &$ \cdot 10^{-3}$ & $0$ && $0.56 $ &$ \cdot 10^{-2}$ & $0.87 $ &$ \cdot 10^{-3}$ & $-0.76 $ &$ \cdot 10^{-4}$\\
 $b$ & $-0.25 $ &$ \cdot 10^{-4}$ & $-0.29 $ &$ \cdot 10^{-4}$ & $0.23 $ &$ \cdot 10^{-5}$ & $0$ &&$-0.15 $ &$ \cdot 10^{-4}$ &$-0.21 $ &$ \cdot 10^{-5}$ &$0.60 $ &$ \cdot 10^{-7}$\\               
\hline
\\[-0.2cm]
 \multicolumn{2}{l}{non-Kramers ions} \\
\hline\hline
 & Pr && Pm && Eu && Tb && Ho && Tm && \\ 
\hline
 $a$ & $0.41 $ &$ \cdot 10^{-5}$ & $0.57 $ &$ \cdot 10^{-5}$ & $0.46 $ &$ \cdot 10^{-5}$ & $\quad0.24 $ &$ \cdot 10^{-2}$ & $0.77 $ &$ \cdot 10^{-3}$ & $-0.57 $ &$ \cdot 10^{-5}$ \\
 $b$ & $0.36 $ &$ \cdot 10^{-6}$ & $0.27 $ &$ \cdot 10^{-7}$ & $0.93 $ &$ \cdot 10^{-9}$ & $0.13 $ &$ \cdot 10^{-5}$ & $0.24 $ &$ \cdot 10^{-5}$ & $0.12 $ &$ \cdot 10^{-7}$ \\
\hline
\end{tabular} 
\caption{Values of the coefficients $a$ and $b$, Eq.~(\ref{p5:eff}), for different rare-earth ions.
We assume that the magnetic field ${\bf B}$ in Eq.~(\ref{p5:eff}) is expressed in units of T (tesla).
 The values of $a$ and $b$ for Kramers ions correspond to the temperature $T=1\ \textrm{K}$.}
\label{p5:table2}
\end{table*}

If we consider the external magnetic field of the order $B \lesssim 1\ \textrm{T (tesla)}$, the system is in the regime 
$\mu_B B \ll kT \ll \Delta \epsilon_{\textrm{cf}} \lesssim \Delta \epsilon_{LS}$.
Numerical analysis of the Hamiltonian (\ref{p5:Hkin}) in this case gives the following 
parametric dependencies for the coefficients $a$ and $b$ describing the effect (\ref{p5:eff}):
\begin{align}
\textrm{non-Kramers ions:} & \quad
a \propto \frac{1}{H_{\textrm{cf}}\,A_{LS}},\ \ b \propto \frac{1}{A_{LS}^2}\ ;\label{p5:nonKr}\displaybreak\\
\textrm{Kramers ions:} & \quad
a \propto \frac{1}{kT\,A_{LS}},\ \ b \propto \frac{H_{\textrm{cf}}}{kT\,A_{LS}^2}\ .\label{p5:Kr}
\end{align}
The above expressions describe variation of the effect for all rare-earth ions, with the  exception of Eu$^{3+}$ and Gd$^{3+}$.
Eu$^{3+}$ ion has the ground state with $J = 0$, so that the first energy interval is determined 
by the fine structure splitting rather than crystal field; in this case the effect includes additional orders of 
$H_{\textrm{cf}}/A_{LS}$: $a \propto \frac{H_{\textrm{cf}}}{A_{LS}^3},
 \ \ b \propto \frac{H_{\textrm{cf}}^3}{A_{LS}^5}$, and so, is suppressed.
The Gd$^{3+}$ ion has a half-filled $4f$-orbital with $L=0$, so the expression 
$\langle [{\bf L} \times {\bf S}] \rangle$, as we have already mentioned, vanishes 
completely for this ion.
Table~\ref{p5:table2} lists the values of $a$ and $b$ for the rare-earth ions Ce through Yb derived from numerical calculations.

The general features of the variation of $a$ and $b$ for different ions can be understood 
with the help of Eqs.~(\ref{p5:nonKr}) and (\ref{p5:Kr}). In particular, the largest values of $a$ 
among the non-Kramers ions correspond to the ions with the smallest first energy interval in 
the crystal field splitting. This agrees well with the formula (\ref{p5:nonKr}). Tb$^{3+}$ and 
Ho$^{3+}$ which have particularly large values of $a$, have very small first energy intervals 
$\sim 5\ \textrm{cm}^{-1}$, whereas the average energy difference in the crystal field 
spectrum is $\sim 100\ \textrm{cm}^{-1}$. Another general tendency observed for all ions is 
that the $a$-effect is larger than the $b$-effect by approximately the factor 
$H_{\textrm{cf}}/A_{LS}$, just as it should be according to Eqs.~(\ref{p5:nonKr}), (\ref{p5:Kr}).
Although the expressions (\ref{p5:nonKr}) and (\ref{p5:Kr}) describe most general trends observed in 
Table~\ref{p5:table2}, there is still considerable dispersion in the values.
This is because the many-electron states of the different ions have significantly different 
structure, and this kind of variation cannot be accounted for in a simple way.

The rare earth ions in RF$_3$ occupy the sites with local $C_2$ symmetry.
In addition, there is the $C_{3,i}$ symmetry which describes different ionic sites in 
the lattice: for any arbitrary site, there are also sites rotated on $2\pi/3$ and $-2\pi/3$ 
(rotation axis is the same as the local $C_2$ axis), as well as the sites with the inverted 
environment.
The local site symmetry describes the geometry of the single site environment: it poses constraints 
on the crystal field parameter values, and hence, on any kinematics that appear locally. 
The existence of local ionic sites with different orientations in the lattice cell and the 
associated symmetry pose limitations on the form of any material (macroscopic) tensor. The kinematical relation (\ref{p5:eff}) corresponds to the macroscopically averaged quantity, and as such, is 
consistent with the combined $C_{6,i}$ symmetry (the local $C_2$, plus $C_{3,i}$ which describes 
the site orientations). The values presented in Table~\ref{p5:table2}, also, correspond to the kinematics averaged over the lattice. 

From equations (\ref{p5:Hat}) and (\ref{p5:eff})
we derive the energy density induced by the weak interaction
\begin{multline}
E_{\textrm{PNC}} = n \langle g | H_{\textrm{PNC}} | g \rangle \\
=a n A E_0Q_W ({\bf E} \cdot [{\bf B}\times{\bf k}])({\bf B}\cdot{\bf k}) \ ,
\label{p5:nEpnc}
\end{multline}
the number density of rare-earth ions in the compound is about
$n\approx 1.8 \cdot 10^{22} \ \textrm{cm}^{-3}$.
In the above equation we neglect the second term on the right-hand side part of Eq.~(\ref{p5:eff}):
although appearance of the pseudovector $b{\bf k}$ is itself an interesting point, in practice,
as it follows from Table~\ref{p5:table2}, the contribution of the $b$-term is negligible
compared to the contribution of the $a$-term.

The macroscopic magnetization corresponding to Eq.~(\ref{p5:nEpnc}) is
\begin{multline}
{\bf M} = -\frac{\partial E_{\textrm{PNC}}}{\partial {\bf B}}
= - n a A E_0 Q_W \Big\{{\bf k}({\bf B}\cdot[{\bf k}\times{\bf E}]) \\
+ [{\bf k}\times{\bf E}]({\bf B}\cdot{\bf k})\Big\} \ .
\label{p5:magn}
\end{multline}
Numerical values of the coefficients $a$ and $A$ are presented in
Tables \ref{p5:table1}, \ref{p5:table2}; we assume that the electric field is expressed in units of
V/cm, and the magnetic field ${\bf B}$ and magnetization ${\bf M}$ are expressed in units of T (tesla).
The temperature is set to $T = 1\ \textrm{K}$, although, this is important for Kramers ions only. 
From Tables \ref{p5:table1}, \ref{p5:table2} and the Eq.~(\ref{p5:magn}) we see that the effect 
is maximal for Dy$^{3+}$, which is a Kramers ion, and Ho$^{3+}$  which is non-Kramers.
Under favorable experimental conditions, it is reasonable to assume the applied magnetic 
field of $B=1\ \textrm{T}$ and the electric field of $E = 10\ \textrm{kV/cm}$. Then,  
the magnetic induction in the sample is
\begin{equation}
\begin{aligned}
\textrm{DyF$_3$:}& \ 4\pi M \sim 0.5 \cdot 10^{-11} \ \textrm{G}\ ,\\[0.1ex]
\textrm{HoF$_3$:}& \ 4\pi M \sim 0.1 \cdot 10^{-11} \ \textrm{G}\ .
\end{aligned}
\end{equation}

Quadratic dependence of $E_{\textrm{PNC}}$ on the value of magnetic field seen in Eq.~(\ref{p5:nEpnc}) is valid only if $\mu_B B \ll kT$ for Kramers ions, and $\mu_B \ll \Delta \epsilon_{\textrm{cf}}$ for non-Kramers ions. For higher values of magnetic field the dependence turns to linear.
This means that for Dy and Ho compounds the PNC magnetization (\ref{p5:magn}) increases linearly with $B$ up to $B\sim 1 \ \textrm{T}$ and then stops to grow.

\section{Discussion}

The parity-violating weak interaction of electrons with the weak charge of rare-earth nuclei
in rare-earth trifluorides leads to the parity nonconserving magnetization presented in
Eq.~(\ref{p5:magn}). The effect depends on the external electric and magnetic fields.
According to our estimates, the effect is maximal in Ho and Dy trifluorides.
Assuming that the external magnetic field is $B = 1\ \textrm{T}$, the external electric field is $E = 10\ \textrm{kV/cm}$, and
the temperature is $T = 1\ \textrm{K}$, we find that the PNC magnetization is about $10^{-11}\ \textrm{G}$.
We can compare this estimate with the current experimental sensitivity that is
under discussion in the literature. According to Ref.~\cite{Lamo02}, it is possible to achieve 
the sensitivity of $3 \cdot 10^{-16}\ \textrm{G}$ for 10 days of averaging with SQUID 
magnetometry, and it even might be possible to do 2 orders of magnitude better with application of different techniques. This leads to conclusion that the value of the PNC magnetization can, at least in principle, be measured to very good precision.

However, an accurate theoretical treatment of the problem with precision of even  several percents does not seem possible. At best, we can expect the accuracy at the level of 20-50\%.
Thus, extraction of the precise value of the weak charge directly from the results of experimental measurements does not seem a viable possibility.
It is possible, however, to look for the variation of the effect in a chain of nuclear isotopes in the same compound. 
In this respect, dysprosium, which has seven naturally occurring isotopes, looks most promising. This opens a possibility for the precise measurements of the Weinberg angle.
There are several issues related to the hypothetical measurements.
The first is the normalization of the effect.
Imagine, there are two samples made of different isotopes. They will be of slightly different shapes and volume, and then one has to perform
precise measurements of the usual electric and magnetic susceptibilities to normalize the
PNC effect per unit volume.
Another issue is related to the possible systematics.  At the moment, we do not see
any macroscopic systematic effects that would imitate the PNC magnetization (\ref{p5:magn}).
Although Eq.~(\ref{p5:magn}), in essence, describes the nonlinear magnetoelectric effect, 
the crucial feature is that it violates parity at fundamental level. 
In the lattice with center of inversion, the usual magnetoelectric effects 
cannot generate this correlation.
A possible source of systematics can be due to the small local imperfections in the lattice which 
destroy the inversion symmetry. However such effects are of local nature and should
average out to zero in the macroscopic sample with about $10^{23}$ rare-earth sites.

Another serious issue (\cite{FoPa90,DePo02}) is the unknown value of nuclear 
neutron 
form factor which appears in the formula for effective nuclear weak charge---the quantity 
that is measured in atomic experiments. Electronic matrix element of the operator 
(\ref{p5:wk1}) of weak charge interaction contains the expression 
$\rho(r) Q_W = \rho(r) (-N + Z(1-4\sin^2\theta_W))$ averaged over the electron radial wave 
functions, $\rho(r)$ here is the nuclear density which replaces the Dirac delta function 
in Eq.~(\ref{p5:wk1}). It is known, however, that the density distribution is different for 
neutrons and protons, so the correct expression to be averaged is 
$(-N\rho_n(r) + Z\rho_p(r)(1-4\sin^2\theta_W))$, where $\rho_n(r)$ and $\rho_p(r)$ 
are the neutron and the proton nuclear density respectively. Thus, in a real experiment one probes 
the effective weak charge
\begin{equation}
\widetilde{Q}_W = -Nq_n + Zq_p(1-4\sin^2\theta_W)\ 
\label{p5:QWeff}
\end{equation}
which contains the neutron and proton form factors $q_n$ and $q_p$.
The form factor $q_p$ can, at least in principle, be measured reliably with
electron scattering, in muonic atoms, etc. 
It is expressed in terms of the proton distribution root-mean-square (RMS) radius $R_p$, 
which is also the RMS radius of nuclear electric charge distribution. 
We assume that the uncertainty in the value of $R_p$ is negligible---the neutron RMS radius $R_n$ and the neutron form factor $q_n$ are known with much lower precision.

Let us estimate the uncertainty in $\sin^2\theta_W$ due to uncertainty in the neutron 
form factor $q_n$. To be specific, let us 
consider the suggested experiment with dysprosium trifluoride.
There are seven stable isotopes of Dy: 156, 158, 160, 161, 162, 163, 164.
The odd and the even isotopes should not be considered together, because the odd isotopes then would give rise to nonmonotonic
dependences due to the presence of the unpaired neutron. Clearly, it is best to analyze the even isotopes only.
This leaves dysprosium-156, 158, 160, 162 and 164---these are completely paired nuclei with practically the same deformations, so the dependence of $q_n$ on the number of neutrons must be smooth and monotonic.

Strictly speaking, in order to find the neutron form factor $q_n$ and the corresponding uncertainty
one has to perform calculations of $q_n$ for deformed nucleus in the rotational
$s$-wave state. However, to estimate the uncertainty, it is sufficient to use the known
formula for spherical nucleus in the ``sharp-edge'' approximation \cite{FoPa90}
\begin{equation}
\label{qn}
q_n = 1- \frac{3}{70} (\alpha Z)^2 \left[ 1+ 5 \left(\frac{R_n}{R_p} 
\right)^2 \right]\ .
\end{equation}
The difference $\Delta R_{np} = R_n - R_p$ was found in Ref.~\cite{TrzciJa01} from 
analysis of experiments with antiprotonic atoms. The empirical fit in Ref.~\cite{TrzciJa01} reads:
\begin{equation}
\Delta R_{np} \,[\textrm{fm}]= (-0.04 \pm 0.03) + (1.01 \pm 0.15)\, \frac{N-Z}{N+Z}\ \ .
\label{p5:nfit}
\end{equation}
According to this expression the uncertainty in $R_n$ is 
$\delta R_n \approx 0.058 \ \textrm{fm}$.
From Eq.(\ref{qn}), uncertainty in the neutron form factor is related to $\delta R_n$ in the following way
\begin{equation}
\label{p5:dqn}
\delta q_n = \frac{3}{7} (\alpha Z)^2\, \frac{R_n}{R^2_p}\, \delta R_n\ .
\end{equation}

The experiment with the chain of isotopes in the same compound would be measuring the ratio 
of $\widetilde{Q}_W$ (\ref{p5:QWeff}) for two different isotopes with $A = Z+N$ and 
$A' = Z+N'$:
\begin{equation}
\label{p5:QWratio}
\frac{\widetilde{Q}_W}{\widetilde{Q}^{\,\prime}_W} = 
\frac{-N q_n + Zq_p(1-4s^2)}{-N' q^{\,\prime}_n + Zq^{\,\prime}_p(1-4s^2)}\ ,
\end{equation}
and thus, it would allow to eliminate the solid-state pre-factor and the associated theoretical uncertainty.
Assuming there is no contribution from experimental error, the uncertainty in 
$s^2 \equiv \sin^2 \theta_W$ that corresponds to the error bounds in the Eq.~(\ref{p5:nfit}) is:
\begin{equation}
\delta s^2 \sim 5 \cdot 10^{-4}\ .
\label{p5:s2uncert}
\end{equation}
Note, that the uncertainties in $q_n$ and $q^{\,\prime}_n$ in Eq.~(\ref{p5:QWratio}) are correlated according to Eqs.~(\ref{p5:dqn}), (\ref{p5:nfit}). Moreover, main contribution to $\delta s^2$ comes from the second term in Eq.~(\ref{p5:nfit}), while contribution of the constant term is negligible.

There is also the experimental uncertainty ${\delta\widetilde{Q}_W}/{\widetilde{Q}_W}$.
In the possible experiment that measures the ratio ${\widetilde{Q}_W}/{\widetilde{Q}^{\,\prime}_W}$, this uncertainty translates into the uncertainty 
$\delta s^2$ in the value of $\sin^2 \theta_W$ according to the formula:
\begin{equation}
\delta s^2 \sim \frac{N^2}{4Z\,\Delta N} \sqrt{2}\,\left(\frac{\delta\widetilde{Q}_W}{\widetilde{Q}_W}\right) = 6\,\left(\frac{\delta\widetilde{Q}_W}{\widetilde{Q}_W}\right) .
\end{equation}
This corresponds to the measurement with two dysprosium isotopes, $^{156}$Dy and $^{164}$Dy, $\Delta N = 8$. In order to match the uncertainty (\ref{p5:s2uncert}), the experiment should be carried out with the precision exceeding ${\delta\widetilde{Q}_W}/{\widetilde{Q}_W} \sim 8 \cdot 10^{-5}$.
According to the estimates, statistical considerations allow for precision of the order $\sim 3\cdot 10^{-5}$, or better, depending on the magnetometry technique.

The current experimental precision for the value of 
$\sin^2 \theta_W$ measured at $Z$-peak, according to Ref.~\cite{PDG}, is $\delta s^2 \sim 1.5 \cdot 10^{-4}$.
The solid-state experiment proposed in the present paper could achieve comparable level of
precision, and, what is more important, would measure the value of the Weinberg angle in the domain of low energies (effective momentum transfer $\sim 30\ \textrm{MeV}$).

The authors would like to thank J. M. Cadogan  and A. I. Milstein for stimulating discussions, and D. Budker and V. A. Dzuba for important comments.


\begin{thebibliography}{99}
\bibitem{PDG} Particle Data Group, Phys. Lett. B {\bf 592}, 1  (2004).
\bibitem{BaZo78} L. M. Barkov and M. S. Zolotorev, JETP Lett., {\bf 27}, 357
(1978); M. J. D. Macpherson, K. P. Zetie, R. B. Warrington, D. N. Stacey, and J. P. Hoare, Phys. Rev. Lett., {\bf 67}, 2784 (1991).
\bibitem{Mee93} D. M. Meekhof, P. Vetter, P. K. Majumder, S. K. Lamoreaux, and E. N. Fortson, Phys. Rev. Lett., {\bf 71}, 3442 (1993).
\bibitem{Edwa95} N. H. Edwards, S. J. Phipp, P. E. G. Baird, and S. Nakayama, Phys. Rev. Lett., {\bf 74}, 2654 (1995); P. A. Vetter, D. M. Meekhof, P. K. Majumder, S. K. Lamoreaux, and E. N. Fortson, Phys. Rev. Lett., {\bf 74},
2658 (1995).
\bibitem{WoBe97}C. S. Wood {\it et al.}, Science {\bf 275}, 1759 (1997).
\bibitem{BeWi99}S. C. Bennett and C. E. Wieman, Phys. Rev. Lett. {\bf 82}, 2484 (1999).
\bibitem{Khriplo91} I. B. Khriplovich, {\it Parity Nonconservation in Atomic Phenomena}, (Gordon \& Breach, 1991).
\bibitem{DzuFla86} V. A. Dzuba, V. V. Flambaum and I. B. Khriplovich, Z. Phys. {\bf D1}, 243 (1986).
\bibitem{DMille95} D. DeMille, Phys. Rev. Lett. {\bf 74}, 4165 (1995).
\bibitem{Shapi68} F. L. Shapiro, Usp. Fiz. Nauk {\bf 95}, 145 (1968), [Sov. Phys. Usp. {\bf 11}, 345 (1968)].
\bibitem{KhriLa97} I. B. Khriplovich and S. K. Lamoreaux, {\it CP Violation Without
Strangeness} (Springer, Berlin, 1997).
\bibitem{VaKo78} B. V. Vasil'ev and E. V. Kolycheva, ZhETF {\bf 74}, 466 (1978), [Sov. Phys. JETP {\bf 47}, 243 (1978)].
\bibitem{Lamo02} S. K. Lamoreaux, Phys. Rev. A {\bf 66}, 022109 (2002).
\bibitem{MuDzu03} T. N. Mukhamedjanov, V. A. Dzuba, and O. P. Sushkov, Phys. Rev. A {\bf 68}, 042103 (2003).
\bibitem{ReCo02} B. C. Regan, E. D. Commins, C. J. Schmidt, and D. DeMille,
Phys. Rev. Lett. {\bf 88}, 071805 (2002).
\bibitem{HeE05} B. J. Heidenreich {\it et al}, physics/0509106.
\bibitem{LiLa04} C.-Y. Liu and S. K. Lamoreaux, Modern Physics Letters A {\bf 19}, 1235 (2004).
\bibitem{MuSu05} T. N. Mukhamedjanov, O. P. Sushkov, and J. M. Cadogan, Phys. Rev. A 71, 012107 (2005).
\bibitem{MuSu05:2} T. N. Mukhamedjanov, O. P. Sushkov, Phys. Rev. A {\bf 72}, 034501 (2005).
\bibitem{BoBo01}M. A. Bouchiat and C. Bouchiat, Eur. Phys. J. D {\bf 15}, 5 (2001).
\bibitem{com1} This is the same averaging that is necessary to proceed from the
single-particle spin-orbit interaction, $a_{ls}{\bf l}\cdot{\bf s}$, to the configuration
spin-orbit interaction $A_{LS}{\bf L}\cdot{\bf S}$.
\bibitem{CFP} D. D. Velkov, {\it Multi-electron coefficients of fractional parentage for the p, d, and f shells}, Ph.D.~thesis at the Johns Hopkins University (2000), unpublished.
\bibitem{MoLe79} C. A. Morrison and R. P. Leavitt, J. Chem. Phys. {\bf 71}, 2366 (1979).
\bibitem{HoGu70} L. Holmes, H.J. Guggenheim, and G.W. Hull, Solid State Commun., {\bf 8}, 2005 (1970).
\bibitem{BleGre88} B. Bleaney {\it et al}, J. Phys. C: Solid State Phys. {\bf 21}, 2721 (1988).
\bibitem{LeLGa84} C. Leycuras, H. LeGall, M. Guillot, and A. Marchand, J. Appl. Phys, {\bf 55}, 
2161 (1984).
\bibitem{FoPa90} E. N. Fortson, Y. Pang, and L. Wilets, Phys. Rev. Lett. {\bf 65}, 2857 (1990).
\bibitem{DePo02} A. Derevianko and S. G. Porsev, Phys. Rev. A {\bf 65}, 052115 (2002). 
\bibitem{TrzciJa01} A. Trzcinska, J. Jastrzebski, P. Lubinski, F. J. Hartmann, R. Schmidt, T. von Egidy, and B. Klos, Phys. Rev. Lett. {\bf 87}, 082501 (2001).


\end{thebibliography}
\end{document}